# 1D Poiseuille Plane Channel Flow by Second Order Response Matrix

Barry D. Ganapol

Department of Aerospace and Mechanical Engineering
University of Arizona
Ganapol@cowboy.ame.arizona.edu

**Abstract.** With increasing miniaturization of diagnostic devices for more effective detection of blood-borne pathogens for example, Poiseuille molecular flow in micro channels has become increasingly relevant. Since continuum mechanics no longer applies for Poiseuille flow when the Knudson number is near or larger than unity, kinetic theory is required to capture the microscopic molecular scattering responsible for channel molecular flow and the velocity profile across a channel. Here, we apply a response matrix solution to the 1D Poiseuille flow with a BGK approximation featuring simplicity with precision by following a consistent numerical formulation leading to high precision, 8-place benchmarks.

## 1. INTRODUCTION

1D plane channel Poiseuille molecular flow, shown in Figure 1, occurs in a micro channel of sufficient width so quantum effects are not relevant but molecular collisions are. The flowing molecules of a gas, driven by a pressure gradient, scatter from the upper and lower walls of the channel whose influence is simulated by $\alpha$, the accommodation coefficient in [0,1], representing diffuse reflection. The fundamental assumptions are a transversely infinite channel with laminar mass molecular flow in the *z*-direction as described by the BGK [1] scattering approximation. Wall specular and diffuse reflection and subsequent molecular scattering within the channel enable a steady state density profile across the channel width (*x*-coordinate) in the *z*-direction. The simplification of the BGK model features local conservation of number density, momentum and energy as well as a non-decreasing variable proportional to entropy. Our goal is to determine the microscopic density perturbation profile from a Maxwellian in order to estimate the macroscopic velocity profile and total channel flow rate in the *z*-direction. In addition to solving the Poiseuille BGK equation, we do so with emphasis on simplicity with precision. This becomes apparent in the analytical form of the solution found in terms of common trigonometric and hyperbolic functions. By generating a sequence of solutions in quadrature order, we estimate the solution for the microscopic velocity profiles at the centerline and exiting the wall, the macroscopic mass velocity and flow rate to be at least eight digits of precision, which is uncommon in the published literature.

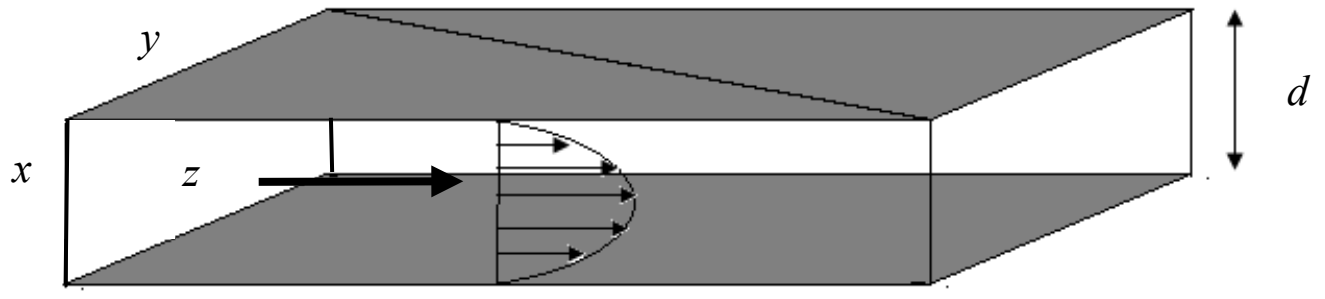

Fig. 1. Poiseuille micro channel flow.

In the next section, we begin with a review of the derivation of the flow equation followed by the corresponding discrete ordinates balance equation. First order forward and backward streams of ODEs appear and couple into a single second order ODE. We choose conveniently normalized hyperbolic solutions for the solution to the diagonalized homogeneous second order form. Finally, a MATLAB program requiring only common matrix inversion and diagonalization provides high-precision numerical results. Where possible, comparison of the response matrix results with literature confirms 5- to 7-place precision.

**1.1 BGK Equation for 1D Poiseuille Plane Channel Flow**

In the following section, we present the derivation of Poiseuille BGK equation. It is brief since an excellent derivation appears in Ref [2]. When linearized, the solution of the nonlinear Boltzmann equation deviates about the Maxwellian $f_0(x,\boldsymbol{c})$ equilibrium distribution by

$$f(x,\boldsymbol{c}) = f_0(x,\boldsymbol{c})\left[1 + h(x,\boldsymbol{c})\right], \tag{1}$$

where the density perturbation $h(x,\boldsymbol{c})$ satisfies the following linearized Boltzmann equation:

$$kc_z + c_x \frac{\partial}{\partial x} h(x,\boldsymbol{c}) = \boldsymbol{L}h(x,\boldsymbol{c}). \tag{2a}$$

$\boldsymbol{L}$ is the BGK operator [1] and

$$k \equiv \frac{1}{P}\frac{\partial P}{\partial x} \tag{2b}$$

is the non-dimensional pressure gradient driving the flow. $c_x$ is the microscopic molecular velocity (aka speed). To specify the perturbed Maxwellian averaged gas velocity in the *z*-direction, we form the following velocity moment by integrating over the *y*- and *z*- directions weighted by all molecular flow velocities $c_z$ in the *z*-direction:

$$Z(x,c_x) = \pi^{-1}\int_{-\infty}^{\infty}\int_{-\infty}^{\infty} e^{-(c_y^2+c_z^2)} c_z h(x,c_x,c_y,c_z)\mathrm{d}c_y\mathrm{d}c_z \ . \tag{3a}$$

One interprets $Z(x,c_x)$ as the perturbed microscopic velocity profile in the *x*-direction of a gas travelling in the *z*-direction. Similarly, integrating Eq(2a) gives

$$\frac{1}{2}k\theta + \theta c_x \frac{\partial}{\partial x} Z(x,c_x) + Z(x,c_x) = \pi^{-1/2}\int_{-\infty}^{\infty} e^{-u^2} Z(x,u)\mathrm{d}u \tag{3b}$$

for $x \in (-d/2, d/2)$, where $\theta$ is the scattering mean free path. Boundary conditions at the walls giving symmetry in *x* are

$$Z(-d/2, c_x) = (1-\alpha)Z(-d/2, -c_x) \tag{3c}$$

$$Z(d/2, -c_x) = (1-\alpha)Z(d/2, c_x)\,, \tag{3d}$$

where $\alpha$ is the accommodation coefficient specifying the degree of wall diffuse and specular scattering. If $\tau \equiv \dfrac{x}{\theta}$, and $\mu \equiv c_x$, then Eq(3b) is

$$\frac{1}{2}k\theta + \mu\frac{\partial}{\partial\tau}Z(\tau,\mu) + Z(\tau,\mu) = \pi^{-1/2}\int_{-\infty}^{\infty} e^{-u^2} Z(\tau,u)\mathrm{d}u \tag{4a}$$

with $\tau \in (-\delta/2, \delta/2)$, $\mu \in (-\infty, \infty)$ and boundary conditions

$$Z(-\delta/2, \mu) = (1-\alpha)Z(-\delta/2, -\mu) \tag{4b}$$

$$Z(\delta/2, -\mu) = (1-\alpha)Z(\delta/2, \mu)\ \ , \tag{4c}$$

where the inverse Knudsen number is

$$\delta \equiv \frac{d}{\theta} \equiv 2a. \tag{4d}$$

Elimination of the inhomogeneous term in Eq(4a) comes after substitution of

$$Z(\tau,\mu) = \frac{1}{2}k\theta[\tau^2 - 2\tau\mu + 2\mu^2 - a^2 - 2Y(\tau,\mu)] \tag{5}$$

to give

$$\left[\mu\frac{\partial}{\partial\tau}+1\right]Y(\tau,\mu) = \int_{-\infty}^{\infty} d\mu'\Psi(\mu')Y(\tau,\mu');\ \mu\in(-\infty,\infty). \tag{6a}$$

with

$$\Psi(\mu) \equiv \frac{e^{-\mu^2}}{\sqrt{\pi}}. \tag{6b}$$

By taking advantage of symmetry about the channel centerline and redefining, $0 \le \tau \le 1$ from the centerline

$$Y(0,\mu) = Y(0,-\mu), \tag{6c}$$

and retaining the lower wall boundary condition for $Y(a,-\mu)$, gives

$$Y(a,-\mu) = (1-\alpha) Y(a,\mu) + g(\mu) \tag{6d}$$

with

$$g(\mu) \equiv \mu\left[\alpha\mu + a(2-\alpha)\right] \tag{6e}$$

for $\mu \in [0,\infty)$. Eq(6a) is a typical transport equation for the "flux distribution" $Y(\tau,\mu)$ similar to neutron transport in a slab except the "angular variable", $\mu$, is not in [-1,1], which is a matter of scaling.

## 2. THEORY OF SECOND ORDER MATRIX SOLUTION

There have been many solutions to Eqs(6). Some of the most successful solutions are found in Refs [2-9], including vibrational methods, elementary singular eigenfunction solutions, Fourier transforms, the FN method and more. Indeed, nearly any method of solving the linear transport equation is feasible including adding and doubling as recently shown [8]. Moreover, the author has developed two response matrix formulations, which are generally more straightforward than previous methods. The first is a first order parity approach that considers the discrete ordinate balance equation as a set of first order ODEs with linearly independent exponential solutions. Decaying exponentials stabilized the solution through an interpretation of the identity [11]. The second casts the solution in a second order The solution form by combining the two directional streams through differentiation similarly found in Ref [12]. This opens up a host of solutions. It is the second order form we pursue here. As will be shown, the second order form provides highly precise benchmarks in contrast to most previous methods cited.

### 2.1. Discrete Ordinates Balance Equation

#### 2.1.1. First Order Form

The fundamental discrete order equations that provide the basis for the solution to the second order form are now derived making liberal use of Ref [8]. By writing Eqs(6) in terms of the new variable $u$ such that $0 \le u \le 1$, and partitioning the scattering integral into forward and backward contributions, Eq(6a) becomes

$$\left[\mu(u)\frac{\partial}{\partial\tau} + 1\right] Y(\tau,\mu(u)) = \int_0^{\infty} d\mu' \Psi(\mu'(u')) \begin{bmatrix} Y(\tau,\mu'(u')) + \\ +Y(\tau,-\mu'(u')) \end{bmatrix}, \tag{7}$$

where the scattering integral splits into two improper integrals. Note the explicit dependence on the new variables $u$ and $u'$, and that $\Psi$ is an even function. With the change of variable, there results

$$Y_0(\tau) \equiv \int_{-\infty}^{\infty} d\mu' \Psi(\mu') Y(\tau,\mu') = \int_0^1 du' \frac{d\mu'(u')}{du'} \Psi(\mu'(u')) \begin{bmatrix} Y(\tau,\mu'(u')) + \\ +Y(\tau,-\mu'(u')) \end{bmatrix} \tag{8}$$

by transforming to $u$. Reference [3] suggests an appropriate choice of $u$ from an infinite number of possibilities as

$$u \equiv e^{-\mu} \tag{9a}$$

to give

$$\mu \equiv -\ln u \tag{9b}$$

$$\frac{d\mu}{du} \equiv -\frac{1}{u} \tag{9b}$$

$$\Psi(\mu) \equiv \frac{1}{\sqrt{\pi}} e^{-\ln(u)^2}. \tag{9c}$$

Therefore, the scattering integral is

$$Y_0(\tau) \equiv \int_{-\infty}^{\infty} d\mu' \Psi(\mu') Y(\tau,\mu') = \int_0^1 du' u'^{-1} \Psi(\mu'(u')) \begin{bmatrix} Y(\tau,\mu'(u')) + \\ +Y(\tau,-\mu'(u')) \end{bmatrix} \tag{10a}$$

enabling approximate half-range integration by Gauss quadrature of order $N$

$$Y_0(\tau) = \sum_{m'=1}^{N} \omega_{m'} \Psi_{m'} \left[ Y_{m'}^{+}(\tau) + Y_{m'}^{-}(\tau) \right] + E_N(\tau), \tag{10b}$$

where

$$Y(\tau, \pm\mu_m) = Y_m^{\pm}(\tau) + \varepsilon_m(\tau) \tag{10c}$$

with the velocity directions shown in Fig. 2. $E_N(\tau)$ is the local quadrature error in $Y_0$ and $\varepsilon_m(\tau)$ is the local quadrature error in the solution. Positive (+) and negative (–) directions are shown in enter in discretized directions through the boundaries at the channel centerline and lower wall and exit out the same boundaries. The discretized directions, $\mu_m$, come from the zeros of the half-range Legendre polynomial of degree $N$

$$P_N(x_m) = 0,\ x_m \equiv 2u_m - 1,\quad m = 1,\ldots,N$$
$$u_m = \frac{1}{2}(1 + x_m) \tag{11a,b}$$

giving the two half-range discrete ordinates for $\mu \in (-\infty, 0)$ and $\mu \in (0, \infty)$

$$\mu_m = -\ln u_m,\ \mu_{N+m} = -\mu_m, \tag{11c}$$

with weights [13]

$$\omega_m = \omega_{N+m} \equiv \frac{2}{(N+1)^2 \left[P_{N+1}(u_m)\right]^2} \frac{1}{u_m}. \tag{11d}$$

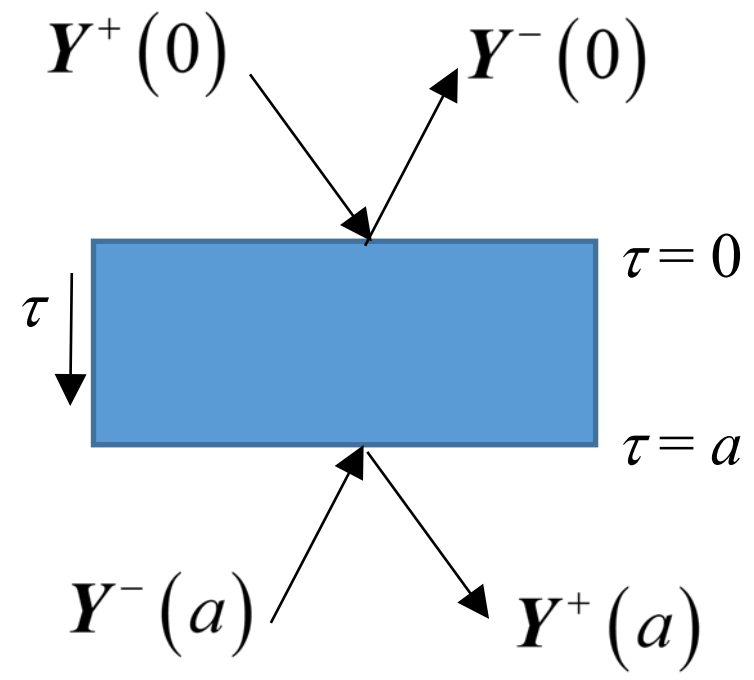

Fig. 2. Molecular directions: + top to bottom/– Bottom to top.

Fig. 2. Now the similarity to discrete ordinates neutron transport in a slab is clear. The channel represents a transmitting and scattering medium, where, molecular density perturbations (particles)

Introducing Eq(10b) into the RHS of Eq(7), evaluating the equation at $\pm\mu_m$ and dropping the error term gives the following discrete ordinates approximation in the positive (+) and negative (–) half-ranges:

$$\left[\mu_m \frac{\partial}{\partial\tau} + 1\right] Y_m^+(\tau) = Y_0(\tau)$$
$$\left[-\mu_m \frac{\partial}{\partial\tau} + 1\right] Y_m^-(\tau) = Y_0(\tau). \tag{12a,b}$$

A more compact form results by defining vectors in the half-ranges

$$Y^\pm(\tau) \equiv \left[Y_1^\pm(\tau) \quad Y_2^\pm(\tau) \quad \ldots \quad Y_N^\pm(\tau)\right]^T. \tag{13}$$

for which Eqs(12) become the following first order vector ODEs:

$$\left[\boldsymbol{I}\frac{d}{d\tau} + \boldsymbol{M}^{-1}\right]\boldsymbol{Y}^+(\tau) = \boldsymbol{M}^{-1}\mathbf{1}Y_0(\tau)$$
$$\left[-\boldsymbol{I}\frac{d}{d\tau} + \boldsymbol{M}^{-1}\right]\boldsymbol{Y}^-(\tau) = \boldsymbol{M}^{-1}\mathbf{1}Y_0(\tau) \tag{14a,b}$$

with

$$
\begin{aligned}
&\boldsymbol{\Psi} \equiv diag\left\{\Psi_m; m=1,...,N\right\} \\
&\boldsymbol{W} \equiv diag\left\{\omega_m; m=1,...,N\right\} \\
&\boldsymbol{M} \equiv diag\left\{\mu_m; m=1,...,N\right\} \\
&\mathbf{1}^T \equiv \left[1_m, m=1,...,N\right]^T = \left[1 \quad 1 \quad ... \quad 1\right]^T .
\end{aligned}
\tag{14c-f}
$$

Note that the unity-vector $[\mathbf{1}]$ makes the RHS a vector of uniform elements. From Eqs(6c,d,e), the boundary conditions are

$$
\begin{aligned}
&Y_{N+m}(0) = Y_m(0) \\
&Y_{N+m}(a) = (1-\alpha)Y_m(a) + g_m \\
&g_m \equiv \alpha\mu_m^2 + a(2-\alpha)\mu_m
\end{aligned}
\tag{15a,b,c}
$$

for $m = 1,\ldots,N$ or as vectors

$$
\begin{aligned}
&\boldsymbol{Y}^+(0) = \boldsymbol{Y}^-(0) \\
&\boldsymbol{Y}^-(a) = (1-\alpha)\boldsymbol{Y}^+(a) + \boldsymbol{g} \\
&\boldsymbol{g} \equiv \alpha\boldsymbol{M}\left[\boldsymbol{M} + a(2-\alpha)\boldsymbol{I}\right]\mathbf{1};
\end{aligned}
\tag{15d,e,f}
$$

and more conveniently in vector form to be used later

$$
\begin{bmatrix} \boldsymbol{Y}^+(0) \\ \boldsymbol{Y}^-(a) \end{bmatrix} = \begin{bmatrix} \mathbf{0} & \boldsymbol{I}_N \\ (1-\alpha)\boldsymbol{I}_N & \mathbf{0} \end{bmatrix} \begin{bmatrix} \boldsymbol{Y}^+(a) \\ \boldsymbol{Y}^-(0) \end{bmatrix} + \begin{bmatrix} \mathbf{0} \\ \boldsymbol{g} \end{bmatrix},
\tag{15g}
$$

where $\boldsymbol{I}_N$ is the matrix identity of size $N$.

### 2.1.2. Equivalent Second Order Form

It is at this point, where the expression of the solution could vary. One has to make a choice on how to solve the set of homogeneous ODEs of Eqs(14a,b). Our choice specifies a unique form of solution with special features. The choice covers any functions that solve a first order vector ODE with constant or variable coefficients. The most notable previous choices have been exponential solutions [3, 5-9], or Fourier transform [10], or others not tried like the Matrix- Raccati formulation [14]-- or the interaction principal [15] to give an invariant embedding solution-- or numerically solving the ODEs by spatial discretization with diamond difference and matrix iteration [16]-- or reforming as the double PN (DPN) moments approximation [17]. All of the methods noted could lead to the response matrix, which we seek here, but we follow a different path.

A second order ODE emerges from adding and subtracting Eqs(14a,b)

$$\frac{d}{d\tau}\left[\boldsymbol{Y}^{+}(\tau)-\boldsymbol{Y}^{-}(\tau)\right]+\boldsymbol{M}^{-1}\left[\boldsymbol{Y}^{+}(\tau)+\boldsymbol{Y}^{-}(\tau)\right]=2\boldsymbol{M}^{-1}\mathbf{1}Y_0(\tau)$$
$$\frac{d}{d\tau}\left[\boldsymbol{Y}^{+}(\tau)+\boldsymbol{Y}^{-}(\tau)\right]+\boldsymbol{M}^{-1}\left[\boldsymbol{Y}^{+}(\tau)-\boldsymbol{Y}^{-}(\tau)\right]=\mathbf{0}, \qquad \text{(16a,b)}$$

which become

$$\frac{d\boldsymbol{\Phi}^{-}(\tau)}{d\tau}+\boldsymbol{M}^{-1}\left[\boldsymbol{\Phi}^{+}(\tau)-2\mathbf{1}Y_0(\tau)\right]=\mathbf{0}$$
$$\frac{d\boldsymbol{\Phi}^{+}(\tau)}{d\tau}+\boldsymbol{M}^{-1}\boldsymbol{\Phi}^{-}(\tau)=\mathbf{0} \qquad \text{(16c,d)}$$

when

$$\boldsymbol{\Phi}^{\pm}(\tau)\equiv\boldsymbol{Y}^{+}(\tau)\pm\boldsymbol{Y}^{-}(\tau). \qquad \text{(16e))}$$

If one differentiates Eq(16d)

$$\frac{d^2\boldsymbol{\Phi}^{+}(\tau)}{d\tau^2}+\boldsymbol{M}^{-1}\frac{d\boldsymbol{\Phi}^{-}(\tau)}{d\tau}=\mathbf{0} \qquad \text{(17a)}$$

and introduces Eq(16c) as

$$\frac{d\boldsymbol{\Phi}^{-}(\tau)}{d\tau}=-\boldsymbol{M}^{-1}\left[\boldsymbol{\Phi}^{+}(\tau)-2\mathbf{1}Y_0(\tau)\right], \qquad \text{(17b)}$$

then a second order ODE emerges

$$\frac{d^2\boldsymbol{\Phi}^{+}(\tau)}{d\tau^2}-\boldsymbol{M}^{-2}\left[\boldsymbol{\Phi}^{+}(\tau)-2\mathbf{1}Y_0(\tau)\right]=\mathbf{0}. \qquad \text{(17c)}$$

The scattering integral $Y_0(\tau)$ will now be expressed in terms of the vector $\boldsymbol{\Phi}^{+}(\tau)$. Since the discretized integral is a dot product on the vector $\boldsymbol{\Phi}^{+}(\tau)$

$$Y_0(\tau)=\sum_{m'=1}^{N}\omega_{m'}\Psi_{m'}\left[Y_{m'}^{+}(\tau)+Y_{m'}^{-}(\tau)\right]=\sum_{m'=1}^{N}\omega_{m'}\Psi_{m'}\Phi_{m'}^{+}(\tau), \qquad \text{(18a)}$$

there results

$$Y_0(\tau)=\boldsymbol{V}^{T}\boldsymbol{\Phi}^{+}(\tau), \qquad \text{(18b)}$$

with

$$\boldsymbol{V}^T \equiv \left[\omega_1\Psi_1 \quad \omega_2\Psi_2 \quad \dots \quad \omega_N\Psi_n\right], \tag{18c}$$

or

$$\boldsymbol{V}^T \equiv \mathbf{1}^T \boldsymbol{C}, \tag{18d}$$

where

$$\boldsymbol{C} \equiv \boldsymbol{W\Psi}; \tag{18e}$$

and therefore

$$Y_0(\tau) = \mathbf{1}^T \boldsymbol{W\Psi\Phi}^+(\tau). \tag{18f}$$

Introducing Eq(18f) into Eq(17c) gives the explicit second order homogeneous equation

$$\frac{d^2\boldsymbol{\Phi}^+(\tau)}{d\tau^2} - \boldsymbol{A}^2\boldsymbol{\Phi}^+(\tau) = \mathbf{0} \tag{19a}$$

with $\boldsymbol{A}^2$

$$\boldsymbol{A}^2 \equiv \boldsymbol{M}^{-2}\left[\boldsymbol{I} - 2\underline{\mathbf{1}}\boldsymbol{C}\right], \tag{19b}$$

squared for convenience. Note that $\underline{\mathbf{1}}$, the uniform one-matrix, is

$$\underline{\mathbf{1}} \equiv \mathbf{1}\mathbf{1}^T = \left\{1_{i,j} = 1;\ i, j = 1, \dots, N\right\}, \tag{20}$$

and the elements of $\boldsymbol{A}^2$ are any real numbers if the quadrature weights are positive.

Once $\boldsymbol{\Phi}^+(\tau)$ is known from the solution to Eq(19a), Eq(16d) gives $\boldsymbol{\Phi}^-(\tau)$ as

$$\boldsymbol{\Phi}^-(\tau) = -\boldsymbol{M}\frac{d\boldsymbol{\Phi}^+(\tau)}{d\tau}. \tag{21}$$

The solution found here for $\boldsymbol{\Phi}^\pm(\tau)$ is but one of many possibilities. For example, a second form comes from differentiating Eq(16c) and then substituting Eq(16d) and the derivative of $Y_0$ from Eq(19f). We would then have another second order ODE but for $\boldsymbol{\Phi}^-(\tau)$. This will be the subject of further research.

## 2.2. Solution of Second Order Form

There are many expressions for the solutions to Eqs(19a) depending on one's knowledge of solutions to ODEs, analytical elegance and computational efficiency. When first encountering a homogeneous vector or scalar second order ODE with constant coefficients, it is most natural to consider exponential solutions or equivalent sines and cosines. These solutions lead to a linear equation for the coefficients of the linear combination of independent solutions after application of boundary conditions. Here, we take a more direct approach.

One anticipates matrix $\boldsymbol{A}$ to have $N$ linearly independent eigenvectors; therefore, $\boldsymbol{A}$ is diagonalizable. Thus,

$$\boldsymbol{A}^2 \equiv \boldsymbol{T}\boldsymbol{\lambda}^2\boldsymbol{T}^{-1} = \boldsymbol{T}diag\left\{\lambda_k^2; k=1,...,N\right\}\boldsymbol{T}^{-1} \tag{22a}$$

or more conveniently by suppressing the $k$ list and letting brackets {} define a diagonal matrix of eigenvalues

$$\boldsymbol{A}^2 = \boldsymbol{T}\left\{\lambda_k^2\right\}\boldsymbol{T}^{-1}, \tag{22b}$$

where $\boldsymbol{T}$ is the matrix of the eigenvectors. The independence of the eigenvectors guarantees the inverse of the $\boldsymbol{T}$ matrix. In addition, since the elements of $\boldsymbol{A}^2$ are all real, the eigenvalues may be imaginary. Since, $\boldsymbol{A}^2 = \boldsymbol{T}\boldsymbol{\lambda}^2\boldsymbol{T}^{-1}$, on substitution into Eq(19a), there results

$$\boldsymbol{T}\boldsymbol{T}^{-1}\frac{d^2\boldsymbol{\Phi}^+(\tau)}{d\tau^2} - \boldsymbol{T}\boldsymbol{\lambda}^2\boldsymbol{T}^{-1}\boldsymbol{\Phi}^+(\tau) = \boldsymbol{0}; \tag{23}$$

with multiplication of the first term by the identity. On defining the reduced vector

$$\boldsymbol{h}(\tau) = \boldsymbol{T}^{-1}\boldsymbol{\Phi}^+(\tau), \tag{24a}$$

Eq(23) becomes

$$\frac{d^2\boldsymbol{h}(\tau)}{d\tau^2} - \boldsymbol{\lambda}^2\boldsymbol{h}(\tau) = \boldsymbol{0}, \tag{24b}$$

which alternatively is a diagonal matrix of scalar ODEs, where the $k^{th}$ second order ODE is

$$\frac{d^2h_k(\tau)}{d\tau^2} - \lambda_k^2 h_k(\tau) = 0; k=1,...,N, \tag{24c}$$

with

$$\boldsymbol{h}(\tau) = diag\left\{h_k(\tau); k=1,...,N\right\}. \tag{24d}$$

The novelty of our approach is rather than assume exponential solutions to the homogeneous $k^{\text{th}}$ ODE, we specify the following linearly independent solutions

$$\begin{aligned} h_k(\tau) &\equiv \frac{\sinh(\lambda_k \tau)}{\sinh(\lambda_k a)} \\ h_k(a-\tau) &\equiv \frac{\sinh(\lambda_k (a-\tau))}{\sinh(\lambda_k a)}, \end{aligned} \tag{25a,b}$$

which are naturally normalization such that

$$h_k(0) = 0, \;\; h_k(a) = 1 \cdot \tag{25c}$$

To check linear independence, the Wronskian is

$$\begin{aligned} W_k(\tau) &\equiv \frac{1}{\sinh(\lambda_k a)} \begin{vmatrix} \sinh(\lambda_k \tau) & \sinh(\lambda_k (a-\tau)) \\ \lambda_k \cosh(\lambda_k \tau) & -\lambda_k \cosh(\lambda_k (a-\tau)) \end{vmatrix} \\ &= -\frac{\lambda_k}{\sinh(\lambda_k a)} \begin{bmatrix} \sinh(\lambda_k \tau)\cosh(\lambda_k (a-\tau)) + \\ +\sinh(\lambda_k (a-\tau))\cosh(\lambda_k \tau) \end{bmatrix} \\ &= -\frac{\lambda_k}{\sinh(\lambda_k a)} \end{aligned} \tag{26}$$

and does not vanish for any eigenvalue including zero, which is not true for simple exponential solutions. The set of solutions of Eqs(25a,b) is particularly convenient since the analytical solution results without the need to solve an additional linear system for the coefficients of the linear combination of the independent solutions as will be shown.

Thus, the general solution to Eq(24b) is

$$\boldsymbol{h}(\tau) = \{h_k(\tau)\}\boldsymbol{C}^+ + \{h_k(a-\tau)\}\boldsymbol{C}^-. \tag{27a}$$

and from Eq(24a),

$$\boldsymbol{\Phi}^+(\tau) = \boldsymbol{T}\boldsymbol{h}(\tau), \tag{27b}$$

and substituting Eq(27a)

$$\boldsymbol{\Phi}^+(\tau) = \boldsymbol{T}\{h_k(\tau)\}\boldsymbol{C}^+ + \boldsymbol{T}\{h_k(a-\tau)\}\boldsymbol{C}^-. \tag{27c}$$

Since, for $\tau = 0$ and $a$

$$\{h_k(0)\} = \mathbf{0},\ \{h_k(a)\} = \boldsymbol{I} \tag{28a}$$

respectively,

$$\begin{aligned} \boldsymbol{C}^+ &= \boldsymbol{T}^{-1}\boldsymbol{\Phi}^+(a) \\ \boldsymbol{C}^- &= \boldsymbol{T}^{-1}\boldsymbol{\Phi}^+(0); \end{aligned} \tag{28b}$$

and the solution from Eq(27c) is

$$\boldsymbol{\Phi}^+(\tau) = \boldsymbol{H}(\tau)\boldsymbol{\Phi}^+(a) + \boldsymbol{H}(a-\tau)\boldsymbol{\Phi}^+(0), \tag{28c}$$

where the matrix function $\boldsymbol{H}(\tau)$ is

$$\boldsymbol{H}(\tau) \equiv \boldsymbol{T}\{h_k(\tau)\}\boldsymbol{T}^{-1} = \boldsymbol{T}\left\{\frac{\sinh(\lambda_k \tau)}{\sinh(\lambda_k a)}\right\}\boldsymbol{T}^{-1}. \tag{28d}$$

Because of the imposed structure of the matrix function $\boldsymbol{H}(\tau)$ satisfies

$$\boldsymbol{H}(0) = 0,\ \boldsymbol{H}(a) = 1, \tag{29}$$

the boundary conditions immediately determine constants $\boldsymbol{C}^+, \boldsymbol{C}^-$ without additional linear algebra, which is a significant advantage of the chosen solutions of Eqs(24c).

In addition, with

$$\begin{aligned} \boldsymbol{H}'(\tau) &= \boldsymbol{T}\left\{\lambda_k \frac{\cosh(\lambda_k \tau)}{\sinh(\lambda_k a)}\right\}\boldsymbol{T}^{-1} \\ \boldsymbol{H}'(a-\tau) &= -\boldsymbol{T}\left\{\lambda_k \frac{\cosh(\lambda_k (a-\tau))}{\sinh(\lambda_k a)}\right\}\boldsymbol{T}^{-1}, \end{aligned} \tag{30a,b}$$

the solution for $\boldsymbol{\Phi}^-(\tau)$ is

$$\boldsymbol{\Phi}^-(\tau) = -\boldsymbol{M}\left[\boldsymbol{H}'(\tau)\boldsymbol{\Phi}^+(a) + \boldsymbol{H}'(a-\tau)\boldsymbol{\Phi}^+(0)\right]. \tag{30c}$$

In principle, with $\boldsymbol{\Phi}^\pm(\tau)$ known, one could find the solution to the first order ODE from Eq(16e); but before we do so, we find the response matrix next.

### 3. The Response Matrix

The response matrix is the matrix, which, when multiplied by the incoming density distributions from the channel centerline, $\boldsymbol{Y}^{+}(0)$, and from accommodation at the bottom wall, $\boldsymbol{Y}^{-}(a)$, (see Fig.2) into the channel, gives the distributions exiting the channel through the centerline, $\boldsymbol{Y}^{-}(0)$, and from the wall, $\boldsymbol{Y}^{+}(a)$. When viewing the solution of Eq(28c) at the boundaries, identities result by construction. Thus, no new information results. However, Eq(30c) for $\boldsymbol{\Phi}^{-}(\tau)$ does give new information such as

$$\begin{aligned} \boldsymbol{\Phi}^{-}(0) &= -\boldsymbol{M}\left[\boldsymbol{H}'(\tau)\big|_0 \boldsymbol{\Phi}^{+}(a) + \boldsymbol{H}'(a-\tau)\big|_0 \boldsymbol{\Phi}^{+}(0)\right] \\ \boldsymbol{\Phi}^{-}(a) &= -\boldsymbol{M}\left[\boldsymbol{H}'(\tau)\big|_a \boldsymbol{\Phi}^{+}(a) + \boldsymbol{H}'(a-\tau)\big|_a \boldsymbol{\Phi}^{+}(0)\right], \end{aligned} \quad \text{(31a,b)}$$

where

$$\begin{aligned} \boldsymbol{H}'(\tau)\big|_0 &= \boldsymbol{T}\left\{\frac{\lambda_k}{\sinh(\lambda_k a)}\right\}\boldsymbol{T}^{-1} \equiv \boldsymbol{B} \\ \boldsymbol{H}'(a-\tau)\big|_0 &= -\boldsymbol{T}\left\{\lambda_k \frac{\cosh(\lambda_k a)}{\sinh(\lambda_k a)}\right\}\boldsymbol{T}^{-1} \equiv \boldsymbol{A} \\ \boldsymbol{H}'(\tau)\big|_a &\equiv -\boldsymbol{A} \\ \boldsymbol{H}'(a-\tau)\big|_a &= -\boldsymbol{B} \end{aligned} \quad \text{(31c-f)}$$

Do not confuse the $\boldsymbol{A}$ in Eqs(31d,e) with the $\boldsymbol{A}$ in Eqs(19). With Eqs(31e-f), Eqs(31a,b) are

$$\begin{aligned} \boldsymbol{\Phi}^{-}(0) &= -\boldsymbol{M}\left[\boldsymbol{B}\boldsymbol{\Phi}^{+}(a) + \boldsymbol{A}\boldsymbol{\Phi}^{+}(0)\right] \\ \boldsymbol{\Phi}^{-}(a) &= \boldsymbol{M}\left[\boldsymbol{A}\boldsymbol{\Phi}^{+}(a) + \boldsymbol{B}\boldsymbol{\Phi}^{+}(0)\right]. \end{aligned} \quad \text{(32a,b)}$$

With the following definitions:

$$\begin{aligned} \boldsymbol{\Phi}^{\pm}(0) &\equiv \boldsymbol{Y}^{+}(0) \pm \boldsymbol{Y}^{-}(0) \\ \boldsymbol{\Phi}^{\pm}(a) &\equiv \boldsymbol{Y}^{+}(a) \pm \boldsymbol{Y}^{-}(a), \end{aligned} \quad \text{(33a,b)}$$

Eqs(32a,b) become

$$\begin{aligned} [\boldsymbol{I}+\boldsymbol{MA}]\boldsymbol{Y}^{+}(0) - [\boldsymbol{I}-\boldsymbol{MA}]\boldsymbol{Y}^{-}(0) &= -\boldsymbol{MB}\left[\boldsymbol{Y}^{+}(a) + \boldsymbol{Y}^{-}(a)\right] \\ [\boldsymbol{I}-\boldsymbol{MA}]\boldsymbol{Y}^{+}(a) - [\boldsymbol{I}+\boldsymbol{MA}]\boldsymbol{Y}^{-}(a) &= \boldsymbol{MB}\left[\boldsymbol{Y}^{+}(0) + \boldsymbol{Y}^{-}(0)\right], \end{aligned} \quad \text{(33c,d)}$$

which, on gathering terms, are

$$\boldsymbol{x}_+\boldsymbol{Y}^+(0)-\boldsymbol{x}_-\boldsymbol{Y}^-(0)=-\boldsymbol{\beta}\left[\boldsymbol{Y}^+(a)+\boldsymbol{Y}^-(a)\right]$$
$$\boldsymbol{x}_-\boldsymbol{Y}^+(a)-\boldsymbol{x}_+\boldsymbol{Y}^-(a)=\boldsymbol{\beta}\left[\boldsymbol{Y}^+(0)+\boldsymbol{Y}^-(0)\right] \tag{33e,f}$$

for

$$\boldsymbol{x}_\pm \equiv \boldsymbol{I} \pm \boldsymbol{MA}$$
$$\boldsymbol{\beta} \equiv \boldsymbol{MB}. \tag{33g,h}$$

A more convenient form of Eqs(33e,f) is the vector equation

$$\begin{bmatrix} \boldsymbol{\beta} & -\boldsymbol{x}_- \\ \boldsymbol{x}_- & -\boldsymbol{\beta} \end{bmatrix}\begin{bmatrix} \boldsymbol{Y}^+(a) \\ \boldsymbol{Y}^-(0) \end{bmatrix}=\begin{bmatrix} -\boldsymbol{x}_+ & -\boldsymbol{\beta} \\ \boldsymbol{\beta} & \boldsymbol{x}_+ \end{bmatrix}\begin{bmatrix} \boldsymbol{Y}^+(0) \\ \boldsymbol{Y}^-(a) \end{bmatrix}. \tag{34}$$

The response matrix (RM) $\boldsymbol{R}$ is therefore

$$\begin{bmatrix} \boldsymbol{Y}^+(a) \\ \boldsymbol{Y}^-(0) \end{bmatrix}=\boldsymbol{R}\begin{bmatrix} \boldsymbol{Y}^+(0) \\ \boldsymbol{Y}^-(a) \end{bmatrix}, \tag{35a}$$

where

$$\boldsymbol{R} \equiv \begin{bmatrix} \boldsymbol{\beta} & -\boldsymbol{x}_- \\ \boldsymbol{x}_- & -\boldsymbol{\beta} \end{bmatrix}^{-1}\begin{bmatrix} -\boldsymbol{x}_+ & -\boldsymbol{\beta} \\ \boldsymbol{\beta} & \boldsymbol{x}_+ \end{bmatrix}. \tag{35b}$$

Note that the vector of entering densities on the RHS of Eq(35a), when multiplied by the response matrix, yields the vector of exiting densities on the left explicitly.

To finish, from the symmetry and boundary conditions given by Eq(15g), Eq(35a) becomes

$$\begin{bmatrix} \boldsymbol{Y}^+(a) \\ \boldsymbol{Y}^-(0) \end{bmatrix}=\boldsymbol{R}\left\{\begin{bmatrix} \boldsymbol{0} & \boldsymbol{I}_N \\ (1-\alpha)\boldsymbol{I}_N & \boldsymbol{0} \end{bmatrix}\begin{bmatrix} \boldsymbol{Y}^+(a) \\ \boldsymbol{Y}^-(0) \end{bmatrix}+\begin{bmatrix} \boldsymbol{0} \\ \boldsymbol{g} \end{bmatrix}\right\} \tag{36a}$$

and solving for the exiting distribution

$$\begin{bmatrix} \boldsymbol{Y}^+(a) \\ \boldsymbol{Y}^-(0) \end{bmatrix}=\left\{\boldsymbol{I}_{2N}-\boldsymbol{R}\begin{bmatrix} \boldsymbol{0} & \boldsymbol{I}_N \\ (1-\alpha)\boldsymbol{I}_N & \boldsymbol{0} \end{bmatrix}\right\}^{-1}\boldsymbol{R}\begin{bmatrix} \boldsymbol{0} \\ \boldsymbol{g} \end{bmatrix}. \tag{36b}$$

From the exiting densities, we infer the interior densities in the next section.

## 4. Solution to Original First Order Form

To convert to the solutions of the original first order ODE of Eqs(14a,b), substitute Eq(16e) into Eqs(28c) and (30c) for $0 \le \tau \le a$

$$\begin{aligned} \boldsymbol{Y}^+(\tau)+\boldsymbol{Y}^-(\tau) &= \boldsymbol{H}(\tau)\left[\boldsymbol{Y}^+(a)+\boldsymbol{Y}^-(a)\right]+\boldsymbol{H}(a-\tau)\left[\boldsymbol{Y}^+(0)+\boldsymbol{Y}^-(0)\right] \\ \boldsymbol{Y}^+(\tau)-\boldsymbol{Y}^-(\tau) &= -\boldsymbol{M}\left\{\boldsymbol{H}'(\tau)\left[\boldsymbol{Y}^+(a)+\boldsymbol{Y}^-(a)\right]+\boldsymbol{H}'(a-\tau)\left[\boldsymbol{Y}^+(0)+\boldsymbol{Y}^-(0)\right]\right\}, \end{aligned} \tag{37a,b}$$

then add and subtract Eqs(37a,b) to give the channel interior microscopic profiles

$$\begin{aligned} \boldsymbol{Y}^+(\tau) &= \frac{1}{2}\left\{\begin{matrix}\left[\boldsymbol{H}(\tau)-\boldsymbol{M}\boldsymbol{H}'(\tau)\right]\left[\boldsymbol{Y}^+(a)+\boldsymbol{Y}^-(a)\right]+ \\ +\left[\boldsymbol{H}(a-\tau)-\boldsymbol{M}\boldsymbol{H}'(a-\tau)\right]\left[\boldsymbol{Y}^+(0)+\boldsymbol{Y}^-(0)\right]\end{matrix}\right\} \\ \boldsymbol{Y}^-(\tau) &= \frac{1}{2}\left\{\begin{matrix}\left[\boldsymbol{H}(\tau)+\boldsymbol{M}\boldsymbol{H}'(\tau)\right]\left[\boldsymbol{Y}^+(a)+\boldsymbol{Y}^-(a)\right]+ \\ +\left[\boldsymbol{H}(a-\tau)+\boldsymbol{M}\boldsymbol{H}'(a-\tau)\right]\left[\boldsymbol{Y}^+(0)+\boldsymbol{Y}^-(0)\right]\end{matrix}\right\}. \end{aligned} \tag{37c,d}$$

Next, introduce the boundary conditions from Eqs(15d,e) to give for $0 \le \tau \le a$

$$\begin{aligned} \boldsymbol{Y}^+(\tau) &= \frac{1}{2}\left\{\begin{matrix}\left[\boldsymbol{H}(\tau)-\boldsymbol{M}\boldsymbol{H}'(\tau)\right]\left[(2-\alpha)\boldsymbol{Y}^+(a)+\boldsymbol{g}\right]+ \\ +2\left[\boldsymbol{H}(a-\tau)-\boldsymbol{M}\boldsymbol{H}'(a-\tau)\right]\boldsymbol{Y}^-(0)\end{matrix}\right\} \\ \boldsymbol{Y}^-(\tau) &= \frac{1}{2}\left\{\begin{matrix}\left[\boldsymbol{H}(\tau)+\boldsymbol{M}\boldsymbol{H}'(\tau)\right]\left[(2-\alpha)\boldsymbol{Y}^+(a)+\boldsymbol{g}\right]+ \\ +2\left[\boldsymbol{H}(a-\tau)+\boldsymbol{M}\boldsymbol{H}'(a-\tau)\right]\boldsymbol{Y}^-(0)\end{matrix}\right\}, \end{aligned} \tag{38a,b}$$

and finally introduce $\boldsymbol{Y}^+(a)$ and $\boldsymbol{Y}^-(0)$ from Eq(36b) in the form

$$\begin{aligned} \boldsymbol{Y}^+(a) &= [\boldsymbol{I}_N \quad \boldsymbol{0}]\left\{\boldsymbol{I}_{2N}-\boldsymbol{R}\begin{bmatrix}\boldsymbol{0} & \boldsymbol{I}_N \\ (1-\alpha)\boldsymbol{I}_N & \boldsymbol{0}\end{bmatrix}\right\}^{-1}\boldsymbol{R}\begin{bmatrix}\boldsymbol{0} \\ \boldsymbol{g}\end{bmatrix} \\ \boldsymbol{Y}^-(0) &= [\boldsymbol{0} \quad \boldsymbol{I}_N]\left\{\boldsymbol{I}_{2N}-\boldsymbol{R}\begin{bmatrix}\boldsymbol{0} & \boldsymbol{I}_N \\ (1-\alpha)\boldsymbol{I}_N & \boldsymbol{0}\end{bmatrix}\right\}^{-1}\boldsymbol{R}\begin{bmatrix}\boldsymbol{0} \\ \boldsymbol{g}\end{bmatrix} \end{aligned} \tag{38c,d}$$

to arrive at the channel interior microscopic velocity distributions in the positive and negative directions

$$
\boldsymbol{Y}^{+}(\tau)=\frac{1}{2}\left\{\begin{array}{l}\left[\boldsymbol{H}(\tau)-\boldsymbol{M}\boldsymbol{H}'(\tau)\right]\boldsymbol{g}+\\ +\left\{\begin{array}{l}\left[\boldsymbol{H}(\tau)-\boldsymbol{M}\boldsymbol{H}'(\tau)\right]\left[(2-\alpha)\left[\boldsymbol{I}_N \quad \boldsymbol{0}\right]\right]+\\ +2\left[\boldsymbol{H}(a-\tau)-\boldsymbol{M}\boldsymbol{H}'(a-\tau)\right]\left[\boldsymbol{0} \quad \boldsymbol{I}_N\right]\end{array}\right\}\left\{\boldsymbol{I}_{2N}-\boldsymbol{R}\begin{bmatrix}\boldsymbol{0} & \boldsymbol{I}_N\\ (1-\alpha)\boldsymbol{I}_N & \boldsymbol{0}\end{bmatrix}\right\}^{-1}\boldsymbol{R}\begin{bmatrix}\boldsymbol{0}\\ \boldsymbol{g}\end{bmatrix}\end{array}\right\}
$$

$$
\boldsymbol{Y}^{-}(\tau)=\frac{1}{2}\left\{\begin{array}{l}\left[\boldsymbol{H}(\tau)+\boldsymbol{M}\boldsymbol{H}'(\tau)\right]\boldsymbol{g}+\\ +\left\{\begin{array}{l}\left[\boldsymbol{H}(\tau)+\boldsymbol{M}\boldsymbol{H}'(\tau)\right]\left[(2-\alpha)\left[\boldsymbol{I}_N \quad \boldsymbol{0}\right]\right]+\\ +2\left[\boldsymbol{H}(a-\tau)+\boldsymbol{M}\boldsymbol{H}'(a-\tau)\right]\left[\boldsymbol{0} \quad \boldsymbol{I}_N\right]\end{array}\right\}\left\{\boldsymbol{I}_{2N}-\boldsymbol{R}\begin{bmatrix}\boldsymbol{0} & \boldsymbol{I}_N\\ (1-\alpha)\boldsymbol{I}_N & \boldsymbol{0}\end{bmatrix}\right\}^{-1}\boldsymbol{R}\begin{bmatrix}\boldsymbol{0}\\ \boldsymbol{g}\end{bmatrix}\end{array}\right\},
$$

(39a,b)

with $\boldsymbol{M}$ from Eq(14e), $\boldsymbol{H}(\tau)$ from Eq(28d), $\boldsymbol{H}'(\tau)$ from Eq(30a), $\boldsymbol{R}$ from Eq(35b) and $\boldsymbol{g}$ from Eq(15f). Thus, in the limit as $N$ approaches infinity, the solution is two infinite dimensional vectors of microscopic velocity profiles specified at $\mu \equiv -\ln u$, where $u$ are the zeros of a half–range Legendre polynomial as its degree approaches infinity.

Eqs(39a,b) are new explicit analytical solutions to the discrete ordinates balance equations [Eqs(14a)]. The only numerical requirements are mapping $\mu$ to the interval [-1,1], matrix diagonalization and matrix inversion. In the next section, we evaluate the macroscopic velocity profile, flow rate and the microscopic velocity profile.

## 5. (Macroscopic) Velocity Profile, Flow Rate and Microscopic Velocity Profile

Expressions for the velocity profile and total flow rate appear in this section. The velocity profile is the Maxwellian weighted moment of the microscopic velocities of Eqs(39); and the flow rate is the integration of the velocity profile over the channel width. The velocity profile is also the RHS (or scattering integral) in Eq(4a).

### 5.1. Velocity Profile

The velocity profile is

$$
q(\tau)\equiv\frac{1}{\sqrt{\pi}k\theta}\int_{-\infty}^{\infty}d\mu' e^{-\mu'^2}Z(\tau,\mu') \qquad (40a)
$$

for $0\le\tau\le a$ and with Eq(5) inserted becomes

$$
\begin{aligned}
q(\tau)&\equiv\frac{1}{2\sqrt{\pi}}\int_{-\infty}^{\infty}d\mu' e^{-\mu'^2}\left[\tau^2-2\tau\mu'-a^2+2\mu'^2-2Y(\tau,\mu')\right]\\
&=\frac{1}{2}\left[\tau^2-a^2+\frac{2}{\sqrt{\pi}}\int_{-\infty}^{\infty}d\mu'\mu'^2 e^{-\mu'^2}\right]-\frac{1}{\sqrt{\pi}}\int_{-\infty}^{\infty}d\mu' e^{-\mu'^2}Y(\tau,\mu')\\
&=\frac{1}{2}\left[1+\tau^2-a^2\right]-Y_0(\tau).
\end{aligned} \qquad (40b)
$$

Introducing $Y_0$ from Eq(18f) yields

$$q(\tau) \equiv \frac{1}{2}\left[1+\tau^2-a^2\right] - \mathbf{1}^T \boldsymbol{W}\boldsymbol{\Psi}\boldsymbol{\Phi}^+(\tau), \tag{41a}$$

and including symmetry and wall conditions from Eq(15d,e) in Eq(28c)

$$\boldsymbol{\Phi}^+(\tau) = \boldsymbol{H}(\tau)\left[(2-\alpha)\boldsymbol{Y}^+(a)+\boldsymbol{g}\right] + 2\boldsymbol{H}(a-\tau)\boldsymbol{Y}^-(0) \tag{41b}$$

to give

$$q(\tau) \equiv \frac{1}{2}\left[1+\tau^2-a^2\right] - \mathbf{1}^T \boldsymbol{W}\boldsymbol{\Psi}\left\{\begin{matrix} \boldsymbol{H}(\tau)\left[(2-\alpha)\boldsymbol{Y}^+(a)+\boldsymbol{g}\right]+ \\ +2\boldsymbol{H}(a-\tau)\boldsymbol{Y}^-(0) \end{matrix}\right\}, \tag{41c}$$

where $\boldsymbol{Y}^+(a)$ and $\boldsymbol{Y}^-(0)$ come from Eqs(38c,d).

**5.2. Flow Rate**

The flow rate is

$$Q(a) \equiv -\frac{1}{2a^2}\int_{-a}^{a} d\tau q(\tau) \tag{42a}$$

and with Eq(40b) becomes

$$Q(a) \equiv -\frac{1}{2a} + \frac{a}{3} + \frac{1}{2a^2}\int_{-a}^{a} d\tau Y_0(\tau). \tag{42b}$$

An analytical result for the integral follows the procedure found in Ref [7]. If we define moments

$$Y_\alpha(\tau) \equiv \int_{-\infty}^{\infty} d\mu \mu^\alpha \Psi(\mu) Y(\tau,\mu), \tag{43a}$$

when the transport equation [Eq(6a)] is multiplied by $\Psi$ and integrated over $\mu$, there results

$$\frac{\partial}{\partial\tau} Y_1(\tau) = 0 \tag{43b}$$

and upon integration

$$Y_1(\tau) = C_1 = Y_1(0) = \int_{-\infty}^{\infty} d\mu\mu\Psi(\mu) Y(0,\mu)$$
$$= \int_{0}^{\infty} d\mu\mu\Psi(\mu)\left[Y(0,\mu) - Y(0,-\mu)\right] = 0. \tag{43c}$$

Now multiply the transport equation [Eq(6a)] again by $\mu\Psi(\mu)$, integrate over $\mu$ to give

$$\frac{\partial}{\partial\tau} Y_2(\tau) + Y_1(\tau) = 0 \tag{44a}$$

and multiply by $\tau$ and integrate over [-*a*,*a*]

$$\int_{-a}^{a} d\tau\tau \frac{\partial}{\partial\tau} Y_2(\tau) + \int_{-a}^{a} d\tau\tau Y_1(\tau) = 0, \tag{44b}$$

which simplifies by integration by parts to

$$\int_{-a}^{a} d\tau Y_2(\tau) = a\left[Y_2(a) + Y_2(-a)\right]. \tag{44c}$$

From symmetry about the centerline,

$$Y_2(a) = Y_2(-a); \tag{45a}$$

therefore,

$$\int_{-a}^{a} d\tau Y_2(\tau) = 2aY_2(a). \tag{45b}$$

Multiply the transport equation [Eq(6a)] by $\mu^2\Psi(\mu)$ and integrate over $\mu$

$$\frac{dY_3(\tau)}{d\tau} + Y_2(\tau) = \frac{1}{2} Y_0(\tau) \tag{46a}$$

and integrate over $\tau$ over [-*a*,*a*]

$$\int_{-a}^{a} d\tau Y_0(\tau) = 2\left(Y_3(a) - Y_3(-a)\right) + 2\int_{-a}^{a} d\tau Y_2(\tau). \tag{46b}$$

Since, from symmetry

$$Y_2(-a) = Y_2(a), \quad Y_3(-a) = -Y_3(a), \tag{47a,b}$$

we find

$$\int_{-a}^{a} d\tau Y_0(\tau) = 4Y_3(a) + 4aY_2(a). \tag{47b}$$

Finally, substituting into Eq(42b) and remembering Eqs(15d,e)

$$Q(a) = -\frac{1}{2a} + \frac{a}{3} + \frac{2}{a^2}\left[Y_3(a) + aY_2(a)\right] \tag{48a}$$

with

$$\begin{aligned}
Y_2(a) &= \int_{-\infty}^{\infty} d\mu\mu^2\Psi(\mu)Y(a,\mu) \\
&= \int_0^{\infty} d\mu\mu^2\Psi(\mu)Y(a,\mu) + \int_0^{\infty} d\mu\mu^2\Psi(\mu)\left[(1-\alpha)Y(a,\mu) + g(\mu)\right] \\
&= (2-\alpha)\int_0^{\infty} d\mu\mu^2\Psi(\mu)Y(a,\mu) + \int_0^{\infty} d\mu\mu^2\Psi(\mu)g(\mu) \\
&= (2-\alpha)\int_0^{\infty} d\mu\mu^2\Psi(\mu)Y(a,\mu) + \frac{3}{8}\alpha + (2-\alpha)\frac{a}{2\sqrt{\pi}}
\end{aligned} \tag{48b}$$

$$\begin{aligned}
Y_3(a) &= \int_{-\infty}^{\infty} d\mu\mu^3\Psi(\mu)Y(a,\mu) \\
&= \int_0^{\infty} d\mu\mu^3\Psi(\mu)Y(a,\mu) - \int_0^{\infty} d\mu\mu^3\Psi(\mu)\left[(1-\alpha)Y(a,\mu) + g(\mu)\right] \\
&= \alpha\int_0^{\infty} d\mu\mu^3\Psi(\mu)Y(a,\mu) - \int_0^{\infty} d\mu\mu^3\Psi(\mu)g(\mu) \\
&= \alpha\int_0^{\infty} d\mu\mu^3\Psi(\mu)Y(a,\mu) - (2-\alpha)\frac{3}{8}a - \frac{\alpha}{\sqrt{\pi}}
\end{aligned} \tag{48c}$$

giving

$$\begin{aligned}
Q(a) = -\frac{1}{2a} + \frac{a}{3} + \frac{2}{a^2}\left[(\alpha-1)\frac{3a}{4}\right] - \frac{1}{\sqrt{\pi}}\left[\alpha - (2-\alpha)\frac{a^2}{2}\right] + \\
+\frac{2}{a^2}\int_0^{\infty} d\mu\mu^2\Psi(\mu)(\mu+a)Y(a,\mu).
\end{aligned} \tag{48d}$$

when Eqs(48b,c) are introduced into Eq(48a). Numerically integrating by Gauss quadrature gives

$$\frac{2}{a^2}\int_0^\infty d\mu\mu^2\Psi(\mu)(\mu+a)Y(a,\mu)\simeq\frac{2}{a^2}\mathbf{1}^T\boldsymbol{W}\boldsymbol{\Psi}\boldsymbol{M}^2(\boldsymbol{M}+a\boldsymbol{I})\boldsymbol{Y}^+(a) \quad . \tag{49}$$

**6. Benchmark Demonstration**

With the analytical solution to the discrete ordinates equations now complete, we turn our attention to the numerical precision of the response matrix method. First, we summarize the response matrix solution and briefly compare its mathematical simplicity to a few previous methods.

The 1D response matrix method based on discrete ordinates discretization of the microscopic molecular velocity profile (direction in neutron transport theory) is a general method of solving transport equations [Eqs(6)] for intensity, flux or density perturbation. After mapping the molecular velocity into the standard interval [-1,1], one solves the resulting set of discretized ODEs [Eqs(14)] over Gauss quadrature by common ODE solution techniques. To the author's knowledge, *all* previous ODE solutions originating from discrete ordinates assume exponential solutions. In his recent publications [11,12] on response matrix solutions however, the author has changed the paradigm. Instead of solving the first order form [Eqs(14)], we construct a second order solution through differentiation [Eq(19a)]. The second order form supports solutions via trigonometric or hyperbolic functions [Eqs(28d)] depending upon whether an eigenvalue of the associated matrix $\boldsymbol{A}$ [Eq(19b)] is real or imaginary both of which occur. Two linearly independent solutions exist even for the zero eigenvalue [Eqs(25a,b)]. With particular construction of the diagonalized solution [Eqs(28d) and (29)], the incoming distribution generates the outgoing distribution through the response matrix [Eqs(35a,b)]. The construction of independent solutions enables a closed form representation of the exiting distributions in terms of density distributions at the boundaries [Eq(36b)]. One finds the final solution when the second order form is algebraically transformed into the first order [Eqs(39)]. The response matrix representation also enables alternative solution representations, which deserve further exploration.

The response matrix solution is a straightforward and readily implemented method of the solution of 1D Poiseuille flow and additional transport equations. Besides round off error from the determination of eigenvalues and eigenvectors, which is probably the most precise of all known numerical methods, the only approximation is the finite number of $N$ discrete ordinates (microscopic velocities). Unlike invariant embedding solutions, there are no X or Y functions requiring iterative solutions. Other methods like Fourier transforms and the FN method require expansion or collocation and integrations for the expansion coefficients. Finally, integral equation solutions require evaluation of the Abramowitz function. In addition, the response matrix applies to heterogeneous channels of separated laminar flow through the multiregion version of the response matrix formulation [8,12].

**6.1 Velocity Profile**

A MATLAB™ program evaluates the velocity profile and flow rate, Eqs(41c) and (48d) respectively. Starting with the velocity profile, Table 1a shows the variation across a channel of half width $a = 1$ for quadrature order $N = 100$. To compensate for the finite number of microscopic velocities allowed, we converge the velocity profile and flow rate in quadrature order $N$. For example, the analytical solution, say for the velocity profile, is actually a limit as the number of discretizations approaches infinity,

$$q(\tau) = \lim_{N\to\infty} q(\tau;N).$$

Since $q(\tau;N)$ exists analytically, we are free to evaluate it at any quadrature order. In this way, one makes a sequence, say in the interval [5,$N$] by increments of 5, and estimates sequence convergence as $N$ approaches infinity. There are many effective choices for assessing sequence convergence. The most common and simplest is element-by-element or sequential convergence

$$q(\tau;5), q(\tau;10), \ldots., q(\tau;5n), \ldots. \tag{50a}$$

where, when the "engineering estimate" of the relative error

$$e(\tau;n) \equiv \left| \frac{q(\tau;n) - q(\tau;n-1)}{q(\tau;n)} \right| \tag{50a}$$

is less that a prescribe error (say $10^{-11}$) the sequence is considered converged. Another effective convergence procedure is non-linear Wynn-epsilon (W-e) convergence acceleration [18]. W-e convergence extrapolates a known sequence to give the next sequence element (estimation of the limit) from previous elements. For our purpose, we use a five-element window passing through the sequence to continually approximate the limit from the last five elements. The algorithm after $L$ elements

$$\begin{aligned}
&\varepsilon_{-1}^{(m)} = 0 \\
&\varepsilon_{0}^{(m)} = s_m,\ m = 0,1,\ldots,L \\
&\varepsilon_{k+1}^{(m)} = \varepsilon_{k-1}^{(m+1)} + \left[\varepsilon_{k}^{(m+1)} - \varepsilon_{k}^{(m)}\right]^{-1}; m = 0,1,\ldots,L-k-1;\ k = 0,1,\ldots,L-1
\end{aligned} \tag{51a}$$

is recursive, where

$$s_m \equiv q(\tau;m+1). \tag{51b}$$

giving following tableau:

$$\begin{array}{cccccc}
\varepsilon_0^{(0)} & \varepsilon_1^{(0)} & \varepsilon_2^{(0)} & \ldots & \varepsilon_{L-1}^{(0)} & \varepsilon_L^{(0)} \\
\varepsilon_0^{(1)} & \varepsilon_1^{(1)} & \varepsilon_2^{(1)} & \ldots & \varepsilon_{L-1}^{(1)} & \\
\varepsilon_0^{(2)} & \ldots & & \ldots & & \\
\ldots & & \varepsilon_2^{(L-2)} & & & \\
 & \varepsilon_1^{(L-1)} & & & & \\
\varepsilon_0^{(L)} & & & & &
\end{array} \tag{51c}$$

where the last value in only the even columns [see arrow] forms the extrapolated sequence giving the relative error

$$e_{We}(\tau;n) \equiv \left| \frac{\varepsilon_n^{(L-n)}(\tau) - \varepsilon_n^{(L-(n-2))}(\tau)}{\varepsilon_n^{(L-n)}(\tau)} \right|. \tag{52}$$

A hybrid form of convergence now applies, where if

$$e_{We}(\tau;n) < e(\tau;n) \tag{53}$$

then $\varepsilon_n^{(L-n)}(\tau)$ is the most precise value; otherwise, $q(\tau;N)$ is the precise value when the inequality is reversed. The values in Table 1a, which is the last element of the iteration sequence 5(5)100, are expected to be precise to better than one unit in the 8th place. This is consistent with the relative errors between the last two iterates given in Table 1b. 38% of the tabular values converged by We acceleration.

Table 1a. Velocity Profile $q(\tau)$ for $a = 1/N = 100$ (CPU= 0.75$s$)

| $\tau \backslash \alpha$ | 0.5 | 0.8 | 0.88 | 0.96 | 1.0 |
|---|---|---|---|---|---|
| 0.0000e+00 | -3.65222151e+00 | -2.31961581e+00 | -2.11740958e+00 | -1.94880072e+00 | -1.87457690e+00 |
| 1.0000e- 01 | -3.64483644e+00 | -2.31214766e+00 | -2.10992110e+00 | -1.94129259e+00 | -1.86705918e+00 |
| 2.0000e- 01 | -3.62257740e+00 | -2.28963849e+00 | -2.08735075e+00 | -1.91866310e+00 | -1.84440085e+00 |
| 3.0000e- 01 | -3.58511714e+00 | -2.25175862e+00 | -2.04936834e+00 | -1.88058161e+00 | -1.80627105e+00 |
| 4.0000e- 01 | -3.53185187e+00 | -2.19790110e+00 | -1.99536633e+00 | -1.82644008e+00 | -1.75206153e+00 |
| 5.0000e- 01 | -3.46178918e+00 | -2.12707160e+00 | -1.92435031e+00 | -1.75524419e+00 | -1.68077806e+00 |
| 6.0000e- 01 | -3.37332081e+00 | -2.03766632e+00 | -1.83471834e+00 | -1.66539407e+00 | -1.59082192e+00 |
| 7.0000e- 01 | -3.26372789e+00 | -1.92699094e+00 | -1.72378372e+00 | -1.55421110e+00 | -1.47951862e+00 |
| 8.0000e- 01 | -3.12791673e+00 | -1.79003943e+00 | -1.58656589e+00 | -1.41674086e+00 | -1.34192713e+00 |
| 9.0000e- 01 | -2.95401978e+00 | -1.61528082e+00 | -1.41162825e+00 | -1.24164283e+00 | -1.16675552e+00 |
| 1.0000e+00 | -2.67640744e+00 | -1.34037200e+00 | -1.13752739e+00 | -9.68381321e- 01 | -8.93924720e- 01 |

Table 1b. Relative error between iterates for $a = 1/N = 100$

| $\tau \backslash \alpha$ | 0.5 | 0.8 | 0.88 | 0.96 | 1.0 |
|---|---|---|---|---|---|
| 0.0000e+00 | 1.2479e-11 | 7.8180e-12 | 7.1166e-12 | 6.5351e-12 | 6.2800e-12 |
| 1.0000e- 01 | 1.1563e-11 | 1.7938e-11 | 1.9106e-11 | 2.0159e-11 | 2.0650e-11 |
| 2.0000e- 01 | 2.3087e-11 | 1.9220e-11 | 1.8736e-11 | 1.8374e-11 | 1.8229e-11 |
| 3.0000e- 01 | 1.5359e-11 | 1.0928e-11 | 1.0290e-11 | 9.7721e-12 | 9.5498e-12 |
| 4.0000e- 01 | 2.9635e-12 | 2.5846e-12 | 3.5318e-12 | 4.3625e-12 | 4.7438e-12 |
| 5.0000e- 01 | 3.8193e-11 | 3.6237e-11 | 3.6299e-11 | 3.6496e-11 | 3.6637e-11 |
| 6.0000e- 01 | 2.4855e-12 | 3.4526e-12 | 4.5183e-12 | 5.4750e-12 | 5.9215e-12 |
| 7.0000e- 01 | 1.5580e-11 | 1.1348e-11 | 1.0775e-11 | 1.0329e-11 | 1.0144e-11 |
| 8.0000e- 01 | 2.4666e-11 | 2.2236e-11 | 2.2220e-11 | 2.2367e-11 | 2.2495e-11 |
| 9.0000e- 01 | 1.7267e-11 | 2.9194e-11 | 3.2238e-11 | 3.5376e-11 | 3.6997e-11 |
| 1.0000e+00 | 1.2712e-11 | 2.5634e-12 | 7.0920e-12 | 6.4186e-12 | 6.1137e-12 |

Curiously, the first eigenvalue is imaginary for the quadrature orders shown in Table 1c.

Table 1c. Occurrence of imaginary eigenvalues ($\lambda_1$)

| $N$ | Re($\lambda$) | Im($\lambda$) | Places |
|---|---|---|---|
| 5 | 0.0000000000e+00 | 4.8253512854e-02 | 0 |
| 10 | 0.0000000000e+00 | 3.5835002846e-03 | 4-5 |
| 25 | 0.0000000000e+00 | 1.9729463267e-05 | 7 |
| 30 | 0.0000000000e+00 | 5.4354269127e-07 | 7-8 |
| 50 | 0.0000000000e+00 | 8.2103403882e-07 | 8 |
| 90 | 0.0000000000e+00 | 1.0041471930e-06 | 8 |

Furthermore, there is no indication in the numerical results of the existence of an imaginary eigenvalue, which previous works did not detected. Table 1c also lists the number of places in agreement for the macroscopic velocities for the quadrature orders.

### 6.2 Flow Rate

Figure 2a displays the flow rate to 8 places for two different methods of calculation over a range of channel widths and accommodation coefficients. The first is for $N = 100$ where 93% of the values converged for the sequence 5(5)100 by W-e acceleration. We find identical values from the adding and doubling (A&D) method of solution [8]. The methods could not be more different, but nevertheless give identical results. While the response matrix relies on eigenvalues and eigenvectors, A&D does not. A&D solves the first order set of ODEs with a Crank-Nicholson approximation in contrast to the response matrix, which is analytic.

Channel widths ($2a$) range from 0.05 to 100. Table 2b shows the corresponding relative error between the two orders $N = 95$ and 100. According to the relative error, the flow rate should be precise to one digit in the last place since the relative error is $10^{-10}$ or less, confirmed in comparison with A&D.

Table 2a. Adding and Doubling and Response Matrix $N = 100^{+}$/CPU = $39s\backslash 1.9s$

| $2a\backslash\alpha$ | 0.5 | 0.8 | 0.88 | 0.96 | 1.0 |
|---|---|---|---|---|---|
| 5.000E-02 | 5.22329643E+00 | 3.08971134E+00 | 2.73834029E+00 | 2.43735442E+00 | 2.30225642E+00 |
| 1.000E-01 | 4.55640624E+00 | 2.70774075E+00 | 2.40604565E+00 | 2.14824142E+00 | 2.03271429E+00 |
| 3.000E-01 | 3.77847230E+00 | 2.24477079E+00 | 2.00106748E+00 | 1.79450880E+00 | 1.70247402E+00 |
| 5.000E-01 | 3.54437089E+00 | 2.10226566E+00 | 1.87662020E+00 | 1.68634239E+00 | 1.60187423E+00 |
| 7.000E-01 | 3.43766932E+00 | 2.03876698E+00 | 1.82201088E+00 | 1.63984952E+00 | 1.55918596E+00 |
| 9.000E-01 | 3.38388693E+00 | 2.00924078E+00 | 1.79763600E+00 | 1.62022302E+00 | 1.54179963E+00 |
| 1.000E+00 | 3.36821820E+00 | 2.00186689E+00 | 1.79205901E+00 | 1.61631243E+00 | 1.53867845E+00 |
| 2.000E+00 | 3.37657376E+00 | 2.04138518E+00 | 1.83856321E+00 | 1.66936555E+00 | 1.59485690E+00 |
| 5.000E+00 | 3.77440185E+00 | 2.43823390E+00 | 2.23505907E+00 | 2.06547805E+00 | 1.99076737E+00 |
| 7.000E+00 | 4.08810781E+00 | 2.74611243E+00 | 2.54143624E+00 | 2.37037511E+00 | 2.29493220E+00 |
| 9.000E+00 | 4.41019024E+00 | 3.06346437E+00 | 2.85756452E+00 | 2.68529504E+00 | 2.60925361E+00 |
| 1.000E+01 | 4.57278306E+00 | 3.22410732E+00 | 3.01770233E+00 | 2.84493372E+00 | 2.76864494E+00 |
| 2.000E+01 | 6.21934716E+00 | 4.85986623E+00 | 4.65065850E+00 | 4.47511900E+00 | 4.39745649E+00 |
| 3.000E+01 | 7.87834275E+00 | 6.51452799E+00 | 6.30419513E+00 | 6.12754285E+00 | 6.04932852E+00 |
| 4.000E+01 | 9.54094965E+00 | 8.17483582E+00 | 7.96390587E+00 | 7.78666297E+00 | 7.70815570E+00 |
| 1.000E+02 | 1.95332586E+01 | 1.81627859E+01 | 1.79507236E+01 | 1.77723604E+01 | 1.76932974E+01 |

+ Choice between linear and Wynn-epsilon convergence

Table 2b. Relative Error between $N = 100$ and 200

| $2a\backslash\alpha$ | 0.5 | 0.8 | 0.88 | 0.96 | 1.0 |
|---|---|---|---|---|---|
| 5.000e-02 | 2.283e-12 | 3.467e-12 | 2.151e-11 | 2.519e-11 | 2.607e-11 |
| 1.000e-01 | 7.913e-12 | 1.281e-12 | 6.296e-13 | 3.890e-12 | 4.586e-12 |
| 3.000e-01 | 5.827e-12 | 1.122e-12 | 4.209e-12 | 6.358e-12 | 6.881e-12 |
| 5.000e-01 | 9.330e-12 | 5.038e-13 | 4.331e-14 | 3.480e-13 | 5.162e-13 |
| 7.000e-01 | 1.582e-12 | 9.273e-12 | 1.517e-13 | 1.850e-12 | 4.409e-13 |
| 9.000e-01 | 1.362e-12 | 9.816e-13 | 2.832e-12 | 7.509e-12 | 6.364e-12 |
| 1.000e+00 | 3.191e-14 | 5.528e-12 | 2.686e-12 | 8.805e-13 | 2.957e-12 |
| 2.000e+00 | 1.255e-13 | 8.223e-14 | 5.217e-12 | 2.658e-12 | 2.564e-12 |
| 5.000e+00 | 8.518e-12 | 4.471e-12 | 6.088e-12 | 7.248e-12 | 7.644e-12 |
| 7.000e+00 | 4.141e-12 | 3.839e-12 | 2.209e-12 | 2.737e-13 | 6.701e-13 |
| 9.000e+00 | 1.610e-12 | 6.941e-12 | 7.240e-12 | 7.310e-12 | 7.291e-12 |
| 1.000e+01 | 5.442e-13 | 5.890e-12 | 6.134e-12 | 6.132e-12 | 6.062e-12 |
| 2.000e+01 | 9.673e-12 | 9.518e-13 | 1.422e-12 | 1.633e-12 | 1.676e-12 |
| 3.000e+01 | 1.049e-12 | 1.644e-12 | 5.702e-13 | 3.856e-14 | 2.048e-13 |
| 4.000e+01 | 1.704e-13 | 5.580e-12 | 2.697e-12 | 1.262e-12 | 8.489e-13 |
| 1.000e+02 | 6.177e-12 | 2.026e-13 | 3.958e-14 | 1.421e-13 | 1.586e-13 |

If we consider smaller channels such as $2a = 0.001$ and 0.0001, shown in Table 2c and as the error observed in Table 2d, the error is of a larger order than for widths above or equal to 0.05. For width 0.001, the flow rate does match four places of the *Fn* method [7] and the error is acceptable, but for width 0.0001, it is not. Below width 0.001; therefore, we conclude the response matrix solution is no longer valid and requires a change in solution strategy, which is the subject of a future investigation. Otherwise, the results of Table 2a agree to all 5 places of Ref [3] with rounding, to all four places of Refs [6 and 7] and to all 7 places of Ref [9] representing some of the most precise values of Poiseuille flow rates found in today's literature.

Table 2c. Convergence of Flow Rate for $N = 5(5)400$, CPU = 23*s*

| $2a\backslash\alpha$ | 0.5 | 0.8 | 0.88 | 0.96 | 1.0 |
|---|---|---|---|---|---|
| 1.0000e-04 | 1.449009e+01 | 7.896153e+00 | 6.844253e+00 | 5.95324e+00 | 5.55660e+00 |
| 1.0000e-03 | 1.068924e+01 | 5.978254e+00 | 5.214431e+00 | 4.56414e+00 | 4.27360e+00 |

Table 2d. Relative Error Relative Error

| $2a\backslash\alpha$ | 0.5 | 0.8 | 0.88 | 0.96 | 1.0 |
|---|---|---|---|---|---|
| 1.000e-04 | 1.719e-04 | 3.136e-04 | 3.608e-04 | 4.133e-04 | 4.417e-04 |
| 1.000e-03 | 7.824e-07 | 1.399e-06 | 1.605e-06 | 1.832e-06 | 1.957e-06 |

For further confirmation, we refer to the interaction principle. The interaction principle considers the channel a homogeneous scattering medium that scatters molecules into and out of the channel boundaries (centerline and wall). The medium equally reflects and transmits molecules entering the channel from either boundary. Thus, if $\boldsymbol{R}_f$ and $\boldsymbol{T}_n$ are the overall reflectance and transmittance matrix coefficients for the channel depending upon only the thickness and scattering properties, one can write from physical reasoning, the exiting density distributions are

$$\boldsymbol{Y}^+(a) = \boldsymbol{T}_n \boldsymbol{Y}^+(0) + \boldsymbol{R}_f \boldsymbol{Y}^-(a)$$
$$\boldsymbol{Y}^-(0) = \boldsymbol{R}_f \boldsymbol{Y}^+(0) + \boldsymbol{T}_n \boldsymbol{Y}^-(a) \qquad (54a,b)$$

or in vector form

$$\begin{bmatrix} \boldsymbol{Y}^+(a) \\ \boldsymbol{Y}^-(0) \end{bmatrix} = \begin{bmatrix} \boldsymbol{T}_n & \boldsymbol{R}_f \\ \boldsymbol{R}_f & \boldsymbol{T}_n \end{bmatrix} \begin{bmatrix} \boldsymbol{Y}^+(0) \\ \boldsymbol{Y}^-(a) \end{bmatrix}, \qquad (54c)$$

which is the interaction principle. Thus, the density exiting the wall results from that reflected from the incoming at the wall through the accommodation coefficient and the contribution transmitted from the centerline. Likewise, the exiting density at the centerline results from reflection of the incoming molecules at the centerline and transmission from the wall to the centerline. In comparing Eq(54c) with Eq(35a), we find

$$\boldsymbol{R} = \begin{bmatrix} \boldsymbol{T}_n & \boldsymbol{R}_f \\ \boldsymbol{R}_f & \boldsymbol{T}_n \end{bmatrix}. \qquad (55)$$

Thus, the response matrix should be block symmetric, which we indeed observed in our calculation.

**6.3 Exiting Microscopic Velocity Profile**

The final calculation is for the exiting microscopic velocity profiles from Eqs(38a,b). Our objective is to find the exiting densities $Y_e$ for an arbitrary series of microscopic velocities $\mu_e$. There are several ways of proceeding. One could solve Eqs(12) with $\mu_m$ replaced by $\mu_e$ for $e = 1,2,\ldots,\varepsilon_L$

$$\left[\mu_e \frac{\partial}{\partial \tau} + 1\right] Y_e^+(\tau) = Y_0(\tau)$$
$$\left[-\mu_e \frac{\partial}{\partial \tau} + 1\right] Y_e^-(\tau) = Y_0(\tau) \qquad (56a,b)$$

requiring additional solutions to ODEs, which are straightforward since we know $Y_0(\tau)$ analytically for each quadrature order and all $\tau$. The simplest approach however is to first note $Y_0(\tau)$ is independent of the microscopic velocity. In this case, one can simply add edits (called faux quadrature edits) to the quadrature list $\mu_m$ as if they were quadratures but have a zero weight [12,19]. In this way, the response matrix formulation interpolates to give the additional edits $\mu_e$ in the output list, but does not change the scattering integral since the edits enter with zero weight. Faux quadratures are also a way to measure the precision of the calculation since they exist in each quadrature order and therefore provide a relative error as the order increases.

The converged exiting distributions, shown in Table 3, are for the following chosen set of faux edits

$$P_{10}\left(x_e\right)=0,\ x_e \equiv 2u_e-1,\quad e=1,\dots,10$$
$$u_e=\frac{1}{2}\left(1+x_e\right) \tag{57a,b}$$
$$\mu_e=-\ln u_e \tag{57c}$$

and for $e = 11$

$$u_{11} = 0.0002$$
$$\mu_{11}=-\ln u_{11}. \tag{57d,e}$$

For only linear convergence, the relative error between quadratures of all orders is $10^{-13}$ and $10^{-14}$ at the centerline and wall respectively. The faux quadrature edits, plotted on the converged solution in Fig. 2, show perfect agreement.

Table 3. Edit exiting microscopic velocity distribution

| $\mu_e$ | $Y^-(0)$ | $Y^+(a)$ |
|---|---|---|
| 1.31325920e-02 | 1.87449029e+00 | 1.43220923e+00 |
| 6.98521513e-02 | 1.87207461e+00 | 1.51665953e+00 |
| 1.74704896e-01 | 1.85316653e+00 | 1.60291016e+00 |
| 3.33101149e-01 | 1.76408706e+00 | 1.67031959e+00 |
| 5.54364553e-01 | 1.65453384e+00 | 1.69443054e+00 |
| 8.54342679e-01 | 1.72997909e+00 | 1.72528197e+00 |
| 1.26124074e+00 | 2.26800624e+00 | 1.97660821e+00 |
| 1.83073806e+00 | 3.74963767e+00 | 2.90547577e+00 |
| 2.69609717e+00 | 7.42638684e+00 | 5.66720498e+00 |
| 4.33921731e+00 | 1.87634259e+01 | 1.52624303e+01 |
| 8.51719319e+00 | 7.22755937e+01 | 6.44636970e+01 |

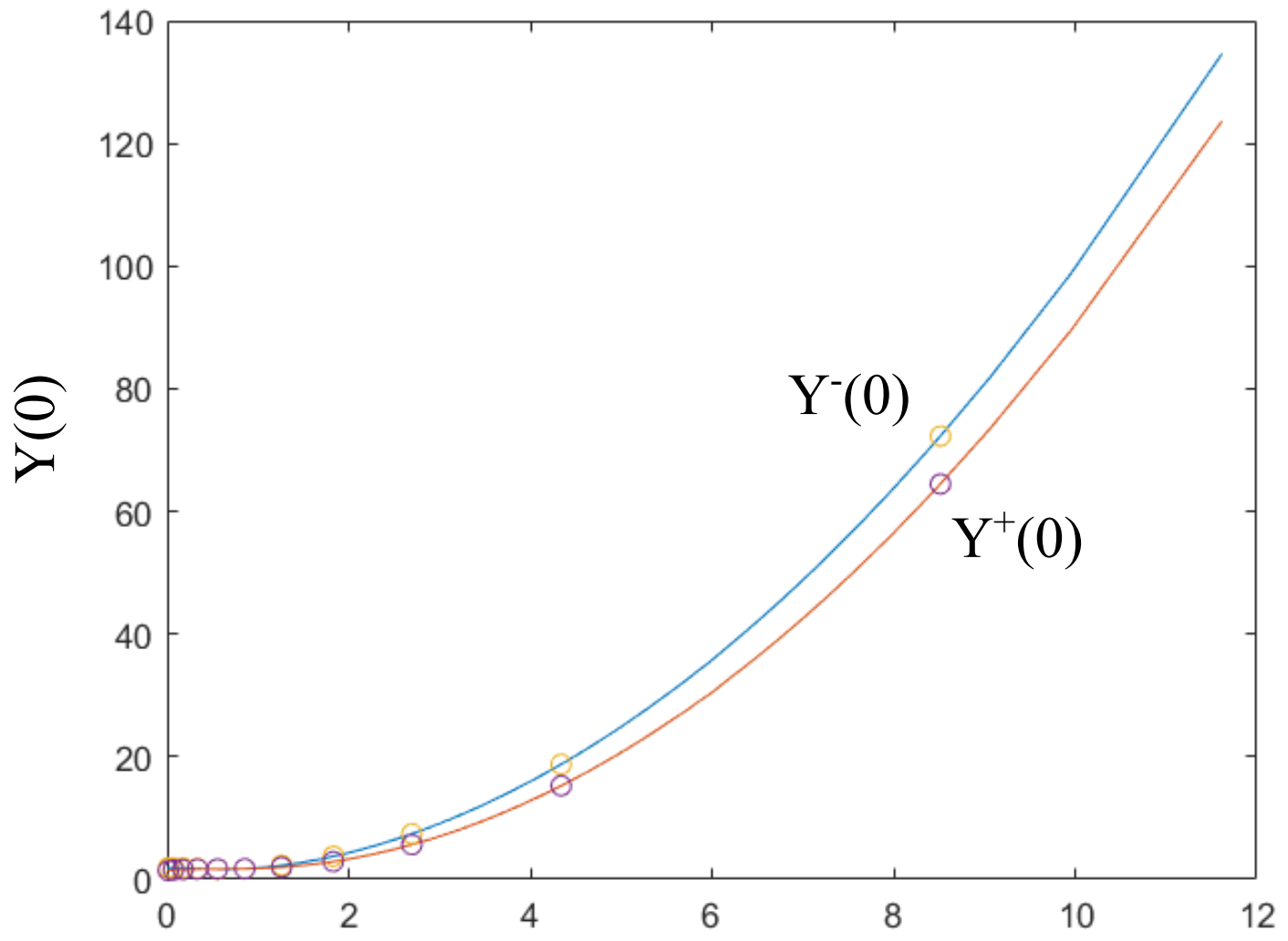

Fig. 1 Faux edits imposed on converged analytical solution.

## 7. CONCLUSION

Poiseuille flow is a fundamental flow of rarefied gas dynamics. Over the years, many excellent solutions have appeared in the literature that are mathematically elegant, but are also mathematically sophisticated and numerically challenging and lack precision. Some give analytical representations at the cost of possibly several time-consuming iterative procedures. However, at the turn of the century, it was the genius of C.E. Siewert and co-workers who revived Chandrasekhar's discrete ordinates method of radiative transfer. It again has became fashionable to solve discretized transport equations, like for Poiseuille flow [3], in terms of practical exponential eigenfunctions and eigenvalues expansions with diagonalization, one of the most developed of all numerical algorithms. These solutions, however, while straightforward, were limited to the set of order *N* first order discrete ordinates ODEs.

In this presentation, we approach the transport solution as second order ODEs. The admitted solutions are trigonometric and hyperbolic functions depending on whether the eigenvalues are real or imaginary. After diagonalization and forming a customized solution, the boundary conditions immediately appear into the general solution without further matrix inversion since there is no need to solve a separate set of linear equations to incorporate boundary conditions. This is but one of many possible solutions representations. As a result, an analytical solution emerges to give the microscopic velocity distribution. By integrating, we specify the macroscopic velocity and the flow rate, which are the common quantities of interest.

By solving the second order form for a series of quadrature orders *N*, one converges the resulting sequence for any dependent variable as *N* become large. Convergence is either linear element by element or through the Wynn-epsilon algorithm. With a converged solution, three benchmarks to 8 places have been established for the macroscopic velocity, flow rate and the microscopic velocity.